\documentclass[fleqn,a4paper,english,12pt,onecolumn]{article}

\usepackage[latin1]{inputenc}
\usepackage{geometry}
\usepackage{graphicx}
\usepackage[intlimits]{amsmath}
\usepackage{amssymb}
\usepackage{babel}
\usepackage[doublespacing]{setspace}
\usepackage{overcite}

\begin{document}

\title{ Electrical Detection of Self-Assembled Polyelectrolyte Multilayers
by a Thin Film Resistor}
\author{Petra A. Neff\,$^{§}$, Ali Naji$^{\dagger }$, Christof Ecker$^{\ddagger }$,\\ 
Bert Nickel$^{\P}$, Regine von Klitzing$^\ddagger $, and Andreas R. Bausch$^\S$\thanks{
To whom correspondence should be addressed: {\texttt{abausch@ph.tum.de}} } \\
\emph{\small $^\S$ Lehrstuhl f\"ur Biophysik - E22, Technische Universit\"at
M\"unchen, Germany}\\
\emph{\small $^\dagger$ Physik Department - T37, Technische Universit\"at
M\"unchen, Germany}\\
\emph{{\small $^{\ddagger }$ Institut für Physikalische Chemie,
Christian-Albrechts-Universität Kiel, Germany}}\\
\emph{\small $^{\P}$ Department für Physik, Ludwig-Maximilians-Universität München, Germany}}
\date{}
\maketitle

\begin{abstract}
The build up of polyelectrolyte multilayers (PEMs) was observed by a
silicon-on-insulator (SOI) based thin film resistor. Differently charged
polyelectrolytes adsorbing to the sensor surface result in defined potential
shifts, which decrease with the number of layers deposited. We model the
response of the device assuming electrostatic screening of polyelectrolyte
charges by mobile ions within the PEMs. The screening length $\kappa ^{-1}$
inside the PEMs was found to be increased compared to the value
corresponding to the bulk solution. Furthermore the partitioning of mobile
ions between the bulk phase and the polyelectrolyte film was employed to
calculate the dielectric constant of the PEMs and the concentration of
mobile charges.
\end{abstract}

\newpage

\textbf{Introduction.} Despite the broad potential applications of
polyelectrolyte multilayers (PEMs) a detailed understanding of the build up
process and the resulting basic physical properties is still elusive. While
the multilayer thickness, the water content, the mechanical properties and
the swelling behavior of different PEMs systems have been extensively
studied, their electrostatic properties are still not fully determined. PEMs
are prepared by the layer--by--layer deposition of polyanions and
polycations from aqueous solutions \cite{deche, bertr}. During the
adsorption process polyanion/polycation complexes are formed with the
previously adsorbed polyelectrolyte layer \cite{farha} leading to a charge
reversal \cite{sukh98}. The exchange of counterions by the oppositely
charged polyelectrolyte could be the reason for the counterion concentration
inside the PEMs to be below the detection limit \cite{schle}. Thus, it seems
that most of the charges within the PEMs are compensated intrinsically by
the opposite polymer charges and not by the presence of small counterions.
Related to the intrinsic charge compensation may be the strong
interdigitation between adjacent layers found by neutron reflectometry 
\cite{schmi, loesc}. While the potential of the outer PEMs surface is well
investigated by electrokinetic measurements \cite{sukh98}, not much is known
about the internal electrostatic properties like ion distribution and
mobility. Using a pH-sensitive fluorescent dye the distribution of protons
within the PEMs has been determined \cite{klitz1}. Assuming Debye screening
and a constant mobility for all ions within the PEMs the potential drop
within polyelectrolyte films composed of poly(allylamine hydrochloride)
(PAH) and poly(sodium 4-styrenesulfonate) (PSS) has been calculated. From
these measurements an independent determination of the ionic strength and
the dielectric constant was not possible. Recent X--ray fluorescence
measurements have been promising in estimating the ion density profile
inside the PEMs giving the total amount of free and condensed ions 
\cite{schol}.

Direct measurements of the potential drop inside the PEMs will be best
suited for determining electrostatic properties such as the Debye length or
the dielectric constant of the PEMs. The capacitance of the PEMs can be
measured by electrochemical methods such as AC voltammetry \cite{slevi}.
Another approach is the use of field effect devices which allows the
determination of the surface potential at the sensor/electrolyte interface.
Obviously, the surface potential variations measured by such a device are
strongly dependent on screening effects inside the adjacent phase. The
deposition of PAH/PSS as well as poly(L-lysine)/DNA multilayers and even DNA
hybridization have been detected by such devices \cite{pouth, fritz, uslu}.
However, a physical model is needed to relate the quantitative response of
the sensors to the dielectric properties and ion mobility inside the PEMs.

Here we show that a silicon-on-insulator (SOI) based thin film resistor is
suited to monitor in real time the build up of polyelectrolyte multilayers
consisting of the strong polyelectrolyte PSS and the weak polyelectrolyte
PAH. The sheet resistance of the field effect device is sensitive to
variations of the potential $\psi _{\mathrm{S}}$ at the silicon oxide
surface. The deposition of the differently charged polyelectrolytes results
in defined potential shifts, which decrease with the number of layers
deposited. Applying a capacitor model, the observed decrease can be
quantitatively explained by assuming reduced electrostatic screening by
mobile charges inside the PEMs compared to the bulk medium outside.

\textbf{Experimental Section.} All chemicals including PSS (MW 70,000) and
PAH (MW 60,000) were purchased from Sigma-Aldrich. Buffers were prepared
using ultrapure water (Millipore, France) with a resistivity $>18$ M$\Omega $
cm. 5 mg/ml polyelectrolyte solutions were prepared by direct dissolution in 
$10$~mM\ Tris buffer at pH 7.5 containing $50$ and $500$~mM\ NaCl,
respectively.

\begin{figure}[h]
\begin{center}
\includegraphics{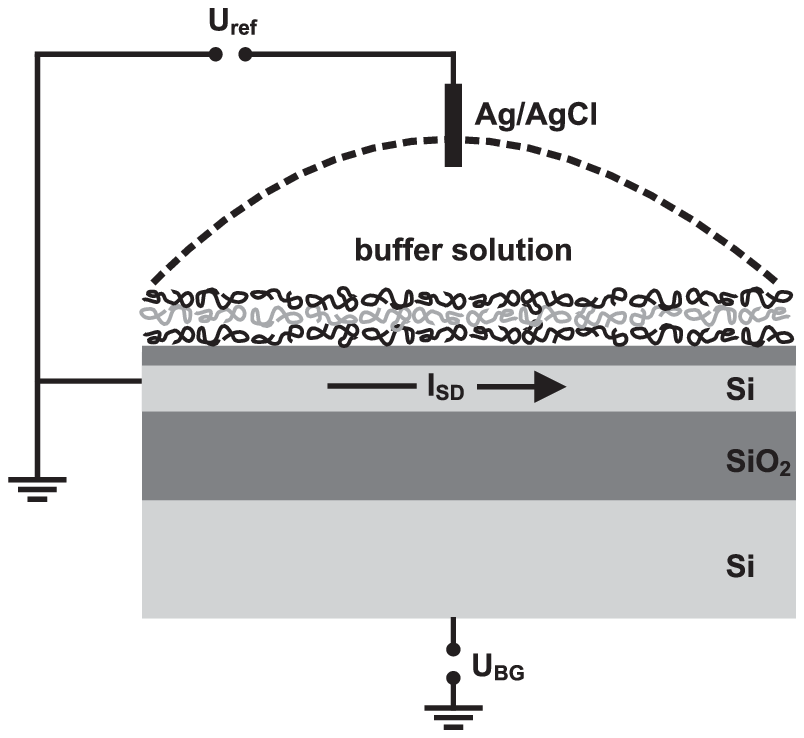}
\end{center}
\caption{Sketch of the setup and the measurement geometry. Silicon is shown
in light grey, silicon oxide in dark grey. From top to bottom: 
native oxide (1-2~nm), conducting top silicon (30~nm), buried oxide (200~nm), 
bulk silicon (675~$\mu$m). 
A voltage is applied between the source and the drain contacts and the resulting 
current $I_{\text{SD}}$ is measured yielding the sheet resistance of the device. 
The carrier concentration in the top silicon layer is tuned by the backgate voltage $U_{\text{BG}}$. 
The potential of the electrolyte solution is controlled by a Ag/AgCl reference electrode. 
A microfluidic device allows the rapid exchange of electrolyte solution.}
\label{setup}
\end{figure}

The sensor chips were fabricated from commercially available
silicon-on-insulator (SOI) wafers (ELTRAN, Canon) using standard
lithographic methods and wet chemical etching as described in detail
elsewhere \cite{nikol1}. The top silicon layer of these wafers was 30 nm
thick and slightly doped with boron ($10^{16}$~cm$^{-3}$). Metal contacts
were deposited in an electron beam evaporation chamber (20~nm Ti, 300~nm
Au). After evaporation, the sensor chips were cleaned using acetone and
isopropanol. The chips were glued into a chip carrier and the contacts were
Au-wire bonded to the carrier. Afterwards the chips were encapsulated with a
silicone rubber to insulate the contacts from the electrolyte solution. The
sheet resistance of the device is dependent on the potential $\psi _{\mathrm{%
S}}$ of the SiO$_{x}$/PEMs interface and was measured as described elsewhere 
\cite{nikol2}. The potential $\psi _{\mathrm{S}}$ was then calculated from
the sheet resistance by a calibration curve of the specific SOI wafer. A
flow chamber was mounted on top of the sensor and a Ag/AgCl reference
electrode was used to control the potential of the electrolyte solution and
for calibration. The setup and the measurement geometry are shown
schematically in Figure \ref{setup}. First, the sensor was equilibrated in
the buffer solution. Next, a calibration measurement was performed as shown elsewhere 
\cite{nikol2}. PAH and PSS solutions were injected twice into the flow
chamber to insure full coverage of the sensor surface, starting with the
positively charged PAH. After obtaining a stable sensor signal, the chamber
was rinsed twice with buffer of the same salt concentration as the
polyelectrolyte solutions. As soon as a stable signal was obtained, the next
polyelectrolyte solution was injected and the procedure was repeated up to
20 times. The sheet resistance of the thin film resistor was monitored
continuously during the multilayer deposition. In separate experiments the
thickness of the deposited polyelectrolyte films was determined by an
ellipsometer (Optrel, Multiscope, Berlin, Germany) in electrolyte solution.
Care was taken to perform the preparation of multilayers as close as
possible to the conditions employed for the deposition on the thin film
resistors. The ellipsometry was carried out in the same buffer solutions
used for the sample preparation.

\begin{figure}[h]
\begin{center}
\includegraphics{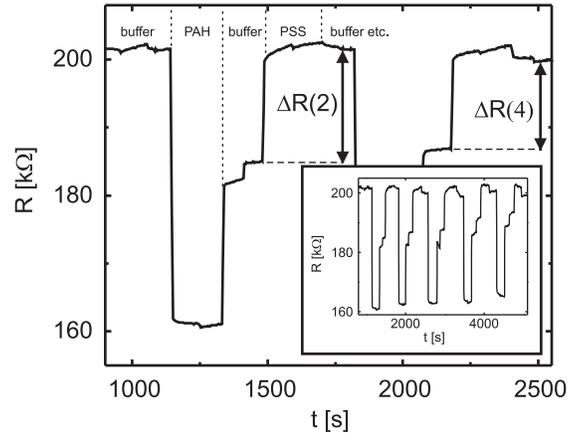}
\end{center}
\caption{The deposition of four PEMs from 50~mM NaCl is shown in detail. The
sheet resistance was monitored continuously during the build-up process. For
each deposition step the polyelectrolyte solution was injected twice. When a
stable signal was obtained the sensor was rinsed twice with buffer. The
resistance change $\Delta R$ between adjacent deposition steps of PAH and
PSS is indicated for two and four adsorbed layers. The potential change
between adjacent deposition steps $\Delta \protect\psi _{\mathrm{S}}$ was
calculated from $\Delta R$ using the calibration data. The inset displays
the subsequent adsorption of 10 layers. }
\label{measurement}
\end{figure}

\begin{figure}[h]
\begin{center}
\includegraphics{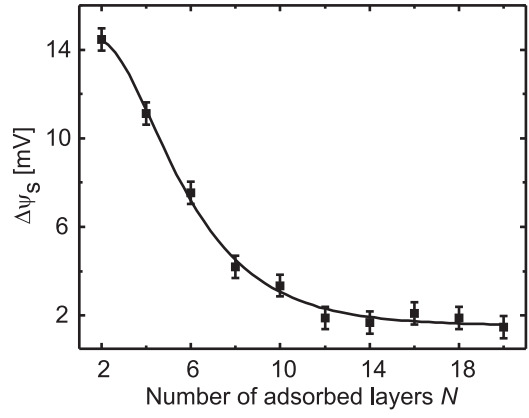}
\end{center}
\caption{The potential change $\Delta\protect\psi _{\text{S}}$ calculated
from the measured change in sheet resistance is plotted as a function of the
number of adsorbed monolayers $N$ for PEMs deposited from 500~mM NaCl. Error
bars are determined from the peak to peak noise of the measurement. Applying
the capacitor model the solid curve was obtained from the fit by Eq. (\protect\ref{delta}).}
\label{deltas_plot}
\end{figure}

\begin{figure}[h]
\begin{center}
\includegraphics{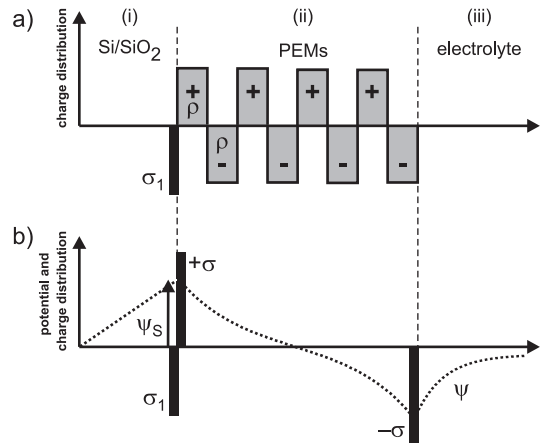}
\end{center}
\caption{The system is modeled using three separate domains: (i) the Si/SiO$%
_{2}$ structure, (ii) the PEMs, and (iii) the electrolyte solution. The
distribution of the immobile charges (the polyelectrolyte charges and the
silicon oxide surface charge $\protect\sigma _{1}$) is shown for two
limiting cases: a) separate layers inside the PEMs with a volume charge
density $\protect\rho $ and b) overlapping layers inside the PEMs with
neutralized charges except for the charges at the sensor surface and the
PEMs/electrolyte interface with the surface charge density of $\pm \frac{ 
\protect\sigma }{2}$. Volume charges $\protect\rho $ are displayed in grey
whereas surface charges $\protect\sigma $ are displayed in black. For case
b) the dotted line schematically shows the potential within the system. The
potential $\protect\psi _{\text{S}}$ at the sensor surface is indicated.}
\label{model}
\end{figure}

\textbf{Results and Discussion.} During the layer-by-layer deposition of the
polycation PAH and the polyanion PSS by alternating buffer exchange, the
sheet resistance of the SOI sensor was continuously observed. The deposition
of the alternately charged polyelectrolytes results in defined responses of
the sensor (Figure \ref{measurement}). The adsorption of PAH decreases the
sheet resistance corresponding to an increased $\psi _{\mathrm{S}}$ (the
potential directly at the SiO$_{x}$/PEMs interface) as positive charges bind
to the surface. The subsequent adsorption of the negatively charged PSS
increases the resistance and thus decreases $\psi _{\mathrm{S}}$. As the SOI
sensor exhibits also a pH sensitivity \cite{nikol2}, the large potential
shift between the PAH deposition and the subsequent washing step can be
attributed to the pH of the PAH solution which is decreased to 6.4 by
dissolution of the week polyelectrolyte in the buffer of pH 7.5. The
deposition of up to 20 monolayers was observable by the field effect device.
The potential change between adjacent deposition steps $\Delta \psi _{%
\mathrm{S}}$ was determined from the measured sheet resistance using the
calibration data and is plotted against the number of adsorbed monolayers as
shown in Figure \ref{deltas_plot}. It can clearly be seen that the potential
jumps $\Delta \psi _{\mathrm{S}}$ decrease with the number of layers
deposited. This is in contrast to electrokinetic studies, where the surface
potential at the outer PEMs/electrolyte interface is measured and the steps
remain constant over a large number of deposited layers \cite{sukh98}. The
observed decrease of $\Delta \psi _{\mathrm{S}}$ can be explained by
adapting a capacitor model \cite{slevi, siu}. The sensor device with the
adsorbed PEMs is modelled assuming three separate domains: (i) The sensor
device, which is modelled as a one dimensional silicon/silicon oxide
structure characterized by its capacitance per area $C_{\mathrm{S}}$. (ii)
The PEMs consisting of $N$ monolayers each of thickness $d$. (iii) The
electrolyte solution outside the PEMs, where a diffuse electrical double
layer is formed at the PEMs/electrolyte interface (Figure \ref{model}). In
order to proceed we need to model the charge distribution within the PEMs.
In general, the layers have a complex charge distribution due to, for
example, interdigitation between polymers from adjacent layers. This
overlapping of layers can even be of the same order as the layer thickness 
\cite{loesc} suggesting charge neutralization within the PEMs \cite{schle}.
Therefore, we consider two limiting cases (Figure \ref{model}): a) separate
layers inside the PEMs with uniform volume charge density of $\rho =\pm 
\frac{\sigma }{d}$, where $\sigma $ may be regarded as the surface charge
density of each layer\footnote{A mesh size of $\approx 30$~nm can be 
estimated for the adsorbed polyelectrolyte layers \cite{netz_joanny}. 
This mesh size is smaller than the estimated effective persistence length of PSS
($\approx 100$~nm), suggesting that alternating polyelectrolyte layers 
form with charge densities which are equal in magnitude.}, and b) overlapping layers such
that polyelectrolyte charges are completely neutralized within the PEMs, and
thus the only uncompensated charges occur at the sensor surface and the
PEMs/electrolyte interface with the surface charge density of $\pm \frac{%
\sigma }{2}$. Electrostatic screening by mobile charges is accounted for
using the linear Debye-Hückel (DH) theory both within the PEMs medium (which
is characterized by the screening length $\kappa ^{-1}$ and the dielectric
constant $\varepsilon $) as well as in the diffuse electrical double layer
of the electrolyte solution (which is characterized by the bulk screening
length $\kappa _{0}^{-1}$ and the dielectric constant for water $\varepsilon
_{\mathrm{w}}$). One can thus identify characteristic Debye capacitances per
area of $C_{\mathrm{P}}=\varepsilon \varepsilon _{0}\kappa $ and $C_{\mathrm{%
D}}=\varepsilon _{\mathrm{w}}\varepsilon _{0}\kappa _{0}$ for the two media,
respectively. The two models a and b considered for the limiting cases
described above yield similar results for the behavior of $\psi _{\mathrm{S}%
} $. On the linear Debye-Hückel level even identical results are obtained,
indicating that the precise charge distribution within the PEMs is not a
critical factor in our model (see Supporting Information). To compare our
measurements with the model predictions, we calculate $\Delta \psi _{\mathrm{%
S}}=\psi _{\mathrm{S}}(N-1)-\psi _{\mathrm{S}}(N)$ which can be simplified
for $\kappa d\ll 1$ and even number of layers, $N$, yielding 
\begin{equation}
\Delta \psi _{\mathrm{S}}(N)=\frac{\sigma \,C_{\mathrm{D}}^{-1}}{\left( C_{%
\mathrm{S}}/C_{\mathrm{P}}+C_{\mathrm{P}}/C_{\mathrm{D}}\right) \sinh \left(
\kappa Nd\right) +\left( 1+C_{\mathrm{S}}/C_{\mathrm{D}}\right) \cosh \left(
\kappa Nd\right) }.  \label{delta}
\end{equation}%
In Eq. (\ref{delta}) the number of layers, $N$, appears only in the
hyperbolic functions. A value for the screening length $\kappa ^{-1}$ inside
the PEMs can be obtained from the measured potential change $\Delta \psi _{%
\mathrm{S}}(N)$ if the thickness $d$ of the polyelectrolyte layers at a
given ionic strength is known. Therefore the monolayer thickness $d$ was
independently measured by ellipsometry. We found an average thickness of $%
d=1.3\pm 0.1$~nm for deposition from 50 mM and $d=2.2\pm 0.1$~nm for
deposition from 500 mM bulk electrolyte solution, respectively. Selected
ellipsometry data have been crosschecked by X--ray synchrotron reflectivity
experiments \cite{reich} confirming the obtained PEMs
thicknesses. As can be seen in Figure \ref{deltas_plot}, the measured $%
\Delta \psi _{\mathrm{S}}$ can be fitted for $\kappa d$ by Eq. (\ref{delta}%
), yielding $\kappa ^{-1}=6.5\pm 1.0$~nm for 500~mM bulk solution. A similar
fit results in $\kappa ^{-1}=6.3\pm 1.0$~nm for the build up of the PEMs at
50~mM bulk solution. Eq. (\ref{delta}) also provides an estimate for the surface 
charge density $\sigma $ of the adsorbed polyelectrolyte layers, assuming that 
$C_{\mathrm{S}}$ is much smaller than $C_{\mathrm{D}}$ 
($1+C_{\mathrm{S}}/C_{\mathrm{D}} \approx 1$). This leads to 
$\sigma = 0.020~\frac{\mathrm{C}}{\mathrm{m}^{2}}$ for the adsorption from 
50~mM and $\sigma = 0.022~\frac{\mathrm{C}}{\mathrm{m}^{2}}$ from 500~mM bulk 
solution.

The relative dielectric constant $\varepsilon $ of the PEMs can be
calculated from $\kappa $, given the definition $\kappa ^{2}=2N_{\mathrm{A}%
}e^{2}c/(\varepsilon \varepsilon _{0}kT)$, where $N_{\mathrm{A}}$ is
Avogadro's number, $e$ is the elementary charge, $c$ is the ion
concentration, and $k$ is the Boltzmann constant. For this, one has to
determine the concentration $c$\ of mobile ions inside the PEMs, which is
set by the thermodynamic equilibrium between the ions in the bulk solution
(of concentration $c_{0}$) and those in the PEMs (see Supporting Information
for details). It turns out that the dominant factor governing the
ion-partitioning in the PEMs/bulk electrolyte system is the Born energy
change, 
\begin{equation}
\Delta \mu =\frac{e^{2}}{8\pi \varepsilon _{0}a}\left[ \frac{1}{\varepsilon }%
-\frac{1}{\varepsilon _{\mathrm{w}}}\right] ,  \label{born}
\end{equation}%
which arises because of the difference in self-energy of the ions (of radius 
$a$) in the PEMs and in the bulk leading to the well-known ion-partitioning
law $c=c_{0}\exp (-\Delta \mu /kT)$ \cite{israe,netz}. Combining the
preceding relations and the definition of $\kappa $, the relative dielectric
constant $\varepsilon $ of the PEMs and the concentration of mobile ions 
can be calculated numerically from the
values obtained for $\kappa $. We find $\varepsilon =30\pm 2$ and $%
\varepsilon =21\pm 1$ for the multilayers adsorbed from 50~mM NaCl and
500~mM NaCl, respectively. The corresponding concentration
of mobile ions inside the PEMs is estimated to be of the order $0.9\pm 0.3$%
~mM and $0.6\pm 0.2$~mM, respectively.

A slightly higher value of $\varepsilon =50\pm 10$ for the relative
dielectric constant of PAH/PSS films has been estimated by comparing pyrene
fluorescence data of the films with that of various isotropic solvents of
low molecular weight \cite{tedes}. Durstock and Rubner \cite{durs01} found
dielectric constants by factor 20 higher for PAH/PSS multilayers in water
vapor. The large deviation from the values of the present paper is not fully
understood. A possible explanation could be differences in swelling
behavior in water vapor and liquid water as shown by neutron reflectometry.

The different values for $\varepsilon $ obtained for PEMs deposited from
different salt concentrations can be interpreted in terms of a different
water content of the polyelectrolyte films. A water content of about 40\%\
was estimated by neutron reflectometry for PAH/PSS films deposited from
different salt concentrations \cite{loesc, stei, wong, carri} suggesting
that the ionic strength of the solution does not change the water content of
the PEMs. Assuming an equal water content for both salt concentrations the
observed variation of the dielectric constant could be ascribed to a
different fraction of immobilized to free water within the polymer layers as
oriented water molecules show a lower dielectric constant. Decreased water
mobility inside PEMs has already been determined by NMR studies \cite{schwa}. 
Comparing measurements at different conditions will be necessary to
further determine the origin of the dielectric constants of PEMs.

Note that the DH approximation used in the present theoretical model is
valid for relatively small electrostatic potentials.
At room temperature, for symmetrical monovalent electrolytes this 
yields an upper limit of 50 mV, which is typically larger than the 
potentials measured in our experiments. In general a full non-linear 
Poisson-Boltzmann analysis would be necessary, which, however, is not 
analytically solvable for the present system. The main advantage of the 
DH approach lies in obtaining a simple analytical expression for the sensor
device functionalized by PEMs enabling a direct comparison with the experimental data.

\textbf{Conclusion.} We were able to show that the recently introduced field
effect device based on SOI is well suited for the quantitative determination
of charge variations at complex interfaces. We apply a capacitor model
including electrostatic screening by mobile charges within the PEMs to
determine their dielectric constant as well as the concentration of mobile
ions inside the polymer film. The origin of the dielectric constants found
for PEMs deposited from different salt concentrations will need to be
addressed further. The presented theoretical description, which is given
here for the PEMs, may prove useful also for the quantitative analysis of
differently functionalized field effect devices.

\textbf{Acknowledgment.} This work was funded by the Deutsche
Forschungsgemeinschaft within the SFB~563 and partially by the
French--German Network and by the Fonds der Chemischen Industrie. The
authors thank Roland Netz for helpful scientific discussions.

\textbf{Supporting Information Available:} Details of the capacitor models
and the thermodynamic equilibrium between the ions in the bulk solution and
the PEMs are given. This material is available free of charge via the
Internet at http://pubs.acs.org.

\newpage

\newpage

\newpage

\newpage

\newpage

\appendix
\section*{Supporting Information}

\subsection*{Theoretical models for the detection of polyelectrolyte multilayers
by the SOI-based thin film resistor}

\subsubsection*{Model a: Alternating layers with volume charges}

We model the polyelectrolyte multilayers sensor system as a series of $N$
alternately positively and negatively charged layers with the volume charge
distribution $\rho =\pm \frac{\sigma }{d}$ (Figure 4a, main text). The
potential in the silicon/silicon oxide structure is assumed to be linear. It
is determined by the capacitance per area of the device $C_{\mathrm{S}}=%
\frac{\varepsilon _{1}}{d_{1}}$ which is given by the effective dielectric
constant $\varepsilon _{1}$ and the thickness $d_{1}$. The charge of the
silicon oxide surface is given by $\sigma _{1}.$ In the PEMs, the Debye-H%
\"{u}ckel (DH) equation%
\begin{equation*}
\frac{\mathrm{d}^{2}}{\mathrm{d}x^{2}}\psi \left( x\right) -\kappa ^{2}\psi
\left( x\right) =-\frac{1}{\varepsilon }\rho \left( x\right) 
\end{equation*}%
is solved for each single polyelectrolyte layer with $\rho \left( x\right)
=\pm \frac{\sigma }{d}$. The screening length inside the PEMs is assumed to
be $\kappa ^{-1}$ and the dielectric constant is $\varepsilon $. The diffuse
electrical double layer in the electrolyte is described by the dielectric
constant for water $\varepsilon _{\mathrm{w}}$ and the screening length $%
\kappa _{0}^{-1}$ which is equal to the effective double layer thickness.
The actual distribution of counterions at the charged surface is diffuse 
and reaches the unperturbed bulk value only at large distances from the 
surface, which is described by an exponential decay of the potential. 
However, a diffuse double layer behaves like a parallel plate capacitor 
in which the separation between the plates is given by the screening length
$\kappa _{0}^{-1}$. Thus we can describe the potential within the electrolyte 
by a plate capacitor with the Debye capacitance per area 
$C_{\mathrm{D}}=\varepsilon _{\mathrm{w}}\varepsilon _{0}\kappa _{0}$.
The surface charges $\sigma _{0}$ and $\sigma _{L}$
represent the space charges within the seminconductor and the electrical
double layer of the electrolyte, respectively. The potential difference
between the bulk semiconductor and the bulk electrolyte is $U_{\mathrm{tot}}$
and is set by the reference electrode. The potential $\psi $ within the $N+2$
domains (i), (ii)$_{1}$, ..., (ii)$_{N},$ and (iii) is given by%
\begin{eqnarray*}
\psi _{\text{(i)}}\left( x\right)  &=&A^{\prime }x+A, \\
\psi _{\text{(ii)}_{n}}\left( x\right)  &=&C_{n}\exp \left( -\kappa x\right)
+D_{n}\exp \left( \kappa x\right) +\psi _{n}\left( x\right) ,\qquad
n=1,\ldots ,N \\
\psi _{\text{(iii)}}\left( x\right)  &=&B^{\prime }x+B,
\end{eqnarray*}%
where $C_{n}\exp \left( -\kappa x\right) +D_{n}\exp \left( \kappa x\right) $
is the general solution of the homogeneous DH\ equation and $\psi _{n}\left(
x\right) $ is a particular solution of the inhomogeneous DH\ equation for $%
\rho \left( x\right) =-\frac{\sigma }{d}\left( -1\right) ^{n}$. We apply the
following boundary conditions%
\begin{eqnarray*}
\psi _{\text{(i)}}\left( 0\right)  &=&0,\qquad \psi _{\text{(i)}}^{\prime
}\left( 0\right) =-\frac{\sigma _{0}}{\varepsilon _{1}}, \\
\psi _{\text{(iii)}}\left( x_{L}\right)  &=&U_{\mathrm{tot}},\qquad \psi _{%
\text{(iii)}}^{\prime }\left( x_{L}\right) =\frac{\sigma _{L}}{\varepsilon _{%
\mathrm{w}}},
\end{eqnarray*}%
\begin{eqnarray*}
\psi _{\text{(ii)}_{1}}\left( x_{1}\right) -\psi _{\text{(i)}}\left(
x_{1}\right)  &=&0,\qquad \varepsilon \psi _{\text{(ii)}_{1}}^{\prime
}\left( x_{1}\right) -\varepsilon _{1}\psi _{\text{(i)}}^{\prime }\left(
x_{1}\right) =-\sigma _{\mathrm{1}}, \\
\psi _{\text{(iii)}}\left( x_{N+1}\right) -\psi _{\text{(ii)}_{N}}\left(
x_{N+1}\right)  &=&0,\qquad \varepsilon _{\mathrm{w}}\psi _{\text{(iii)}%
}^{\prime }\left( x_{N+1}\right) -\varepsilon \psi _{\text{(ii)}%
_{N}}^{\prime }\left( x_{N+1}\right) =0, \\
\psi _{\text{(ii)}_{n}}\left( x_{n}\right) -\psi _{\text{(ii)}_{n-1}}\left(
x_{n}\right)  &=&0,\qquad \varepsilon \psi _{\text{(ii)}_{n}}^{\prime
}\left( x_{n}\right) -\varepsilon \psi _{\text{(ii)}_{n-1}}^{\prime }\left(
x_{n}\right) =0,\qquad n=2,\ldots ,N
\end{eqnarray*}%
with%
\begin{eqnarray*}
x_{n} &=&d_{1}+\left( n-1\right) d,\qquad n=1,\ldots ,N+1, \\
x_{L} &=&d_{1}+Nd+\kappa _{0}^{-1}.
\end{eqnarray*}%
These conditions allow us to calculate the potential $\psi _{\mathrm{S}%
}=\psi (x_{1})$ which determines the sheet resistance of the device. If we
define the Debye capacitance per area $C_{\mathrm{P}}=\varepsilon
\varepsilon _{0}\kappa $ within the polyelectrolyte medium we can write $%
\psi _{\mathrm{S}}$ as%
\begin{equation*}
\psi _{\mathrm{S}}(N)=\frac{\left( \sigma _{\mathrm{1}}+\sigma _{\mathrm{eff}%
}\right) \left[ \frac{1}{C_{\mathrm{P}}}\sinh \left( \kappa Nd\right) +\frac{%
1}{C_{\mathrm{D}}}\cosh \left( \kappa Nd\right) \right] +\left( U_{\mathrm{%
tot}}-\frac{1}{C_{\mathrm{D}}}(-1)^{N}\sigma _{\mathrm{eff}}\right) }{\left(
C_{\mathrm{S}}/C_{\mathrm{P}}+C_{\mathrm{P}}/C_{\mathrm{D}}\right) \sinh
\left( \kappa Nd\right) +\left( 1+C_{\mathrm{S}}/C_{\mathrm{D}}\right) \cosh
\left( \kappa Nd\right) }
\end{equation*}%
with the effective polyelectrolyte surface charge 
\begin{equation*}
\sigma _{\mathrm{eff}}=\frac{1}{\kappa d}\left[ \frac{1-\exp \left( -\kappa
d\right) }{1+\exp \left( -\kappa d\right) }\right] \sigma .
\end{equation*}

\subsubsection*{Model b: Charges overlapping within the PEMs}

The polyelectrolyte multilayers are modeled as overlapping and charges are
assumed to be completely neutralized within the PEMs. Thus the only
uncompensated charges occur at the sensor surface and the PEMs/electrolyte
interface (Figure 4b, main text). In this case the potential within the
three domains is given by%
\begin{eqnarray*}
\psi _{\text{(i)}}\left( x\right) &=&A^{\prime }x+A, \\
\psi _{\text{(ii)}}\left( x\right) &=&C\exp \left( -\kappa x\right) +D\exp
\left( \kappa x\right) , \\
\psi _{\text{(iii)}}\left( x\right) &=&B^{\prime }x+B.
\end{eqnarray*}%
We apply the following boundary conditions%
\begin{eqnarray*}
\psi _{\text{(i)}}\left( 0\right) &=&0,\qquad \psi _{\text{(i)}}^{\prime
}\left( 0\right) =-\frac{\sigma _{0}}{\varepsilon _{1}}, \\
\psi _{\text{(iii)}}\left( x_{L}\right) &=&U_{\mathrm{tot}},\qquad \psi _{%
\text{(iii)}}^{\prime }\left( x_{L}\right) =\frac{\sigma _{L}}{\varepsilon _{%
\mathrm{w}}},
\end{eqnarray*}%
\begin{eqnarray*}
\psi _{\text{(ii)}}\left( x_{1}\right) -\psi _{\text{(i)}}\left(
x_{1}\right) &=&0,\qquad \varepsilon \psi _{\text{(ii)}}^{\prime }\left(
x_{1}\right) -\varepsilon _{1}\psi _{\text{(i)}}^{\prime }\left(
x_{1}\right) =-\left( \sigma _{1}+\frac{\sigma }{2}\right) , \\
\psi _{\text{(iii)}}\left( x_{2}\right) -\psi _{\text{(ii)}}\left(
x_{2}\right) &=&0,\qquad \varepsilon _{\mathrm{w}}\psi _{\text{(iii)}%
}^{\prime }\left( x_{2}\right) -\varepsilon \psi _{\text{(ii)}}^{\prime
}\left( x_{2}\right) =\left( -1\right) ^{N}\frac{\sigma }{2},
\end{eqnarray*}%
with%
\begin{eqnarray*}
x_{1} &=&d_{1}, \\
x_{2} &=&d_{1}+Nd, \\
x_{L} &=&d_{1}+Nd+\kappa _{0}^{-1}.
\end{eqnarray*}

\begin{figure}[h!]
\begin{center}
\includegraphics{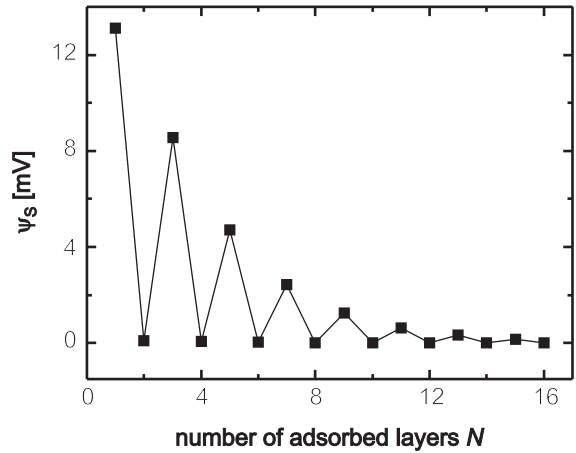}
\end{center}
\caption{The potential $\protect\psi _{\text{S}}$ was calculated within the
model for the deposition from a salt concentration of 500~mM\ and plotted as
a function of the number of adsorbed layers $N$. The solid lines are a guide
to the eye. For the calculation the following parameters were assumed: $C_{%
\text{S}}=2\times 10^{-3}~\frac{\text{F}}{\text{m}^{2}}$, $d=2.2$~nm, $%
\protect\kappa ^{-1}=6.5$~nm, $\protect\varepsilon =21$, $\protect\sigma %
_{1}=-1.25\times 10^{-2}~\frac{\text{C}}{\text{m}^{2}}$ and $\protect\sigma %
=2.5\times 10^{-2}~\frac{\text{C}}{\text{m}^{2}}$. In the calculation the
voltage between the bulk semiconductor and the bulk electrolyte was set to $7
$~mV.}
\label{calculation}
\end{figure}

These conditions allow us to calculate the potential $\psi _{\mathrm{S}%
}=\psi (x_{1})$ and we can write $\psi _{\mathrm{S}}$ according to%
\begin{equation*}
\psi _{\mathrm{S}}(N)=\frac{\left( \sigma _{\mathrm{1}}+\frac{\sigma }{2}%
\right) \left[ \frac{1}{C_{\mathrm{P}}}\sinh \left( \kappa Nd\right) +\frac{1%
}{C_{\mathrm{D}}}\cosh \left( \kappa Nd\right) \right] +\left( U_{\mathrm{tot%
}}-\frac{1}{C_{\mathrm{D}}}(-1)^{N}\frac{\sigma }{2}\right) }{\left( C_{%
\mathrm{S}}/C_{\mathrm{P}}+C_{\mathrm{P}}/C_{\mathrm{D}}\right) \sinh \left(
\kappa Nd\right) +\left( 1+C_{\mathrm{S}}/C_{\mathrm{D}}\right) \cosh \left(
\kappa Nd\right) }.
\end{equation*}%
This is the same result which was obtained for model a if $\sigma _{\mathrm{%
eff}}$ in that model is replaced by $\frac{\sigma }{2}$. The two models
represent limiting cases of the charge distribution. Thus in our models the
charge distribution within the PEMs is not crucial for the values obtained
for $\kappa ^{-1}$ and $\varepsilon $, respectively. For $\kappa d\ll 1$ and
even numbers of $N$ the potential difference $\Delta \psi _{\mathrm{S}}=\psi
_{\mathrm{S}}(N-1)-\psi _{\mathrm{S}}(N)$ simplifies to eq 1 (main text).
The potential $\psi _{\mathrm{S}}$ can now be calculated within the present
model as a function of the number $N$ of layers deposited as shown in Figure %
\ref{calculation}.

\subsection*{Ion-partitioning between PEMs and the bulk solution}

One can calculate the concentration of ions inside the PEMs from the
thermodynamic equilibrium condition with the bulk solution by minimizing the
total free energy density (that is free energy per unit volume) of the
PEMs/bulk electrolyte system as follows.

The total free energy density consists of the free energy density of the
electrical double layer outside the PEMs, ${\mathcal{F}}_{0}$, and the free
energy density of the PEMs medium ${\mathcal{F}}_{\mathrm{P}}$. Within the
Debye-H\"{u}ckel description, the former contribution may be written as ${%
\mathcal{F}}_{0}={\mathcal{F}}_{0}^{\mathrm{MF}}+{\mathcal{F}}_{0}^{\mathrm{%
corr}}+{\mathcal{F}}_{0}^{\mathrm{self}}$ (in units of $kT$) comprising the
mean-field electrostatic free energy density of the double-layer, ${\mathcal{%
F}}_{0}^{\mathrm{MF}}$, the Debye-H\"{u}ckel correlation (or excess) free
energy density, ${\mathcal{F}}_{0}^{\mathrm{corr}}$, and the self-energy
density of the ions, ${\mathcal{F}}_{0}^{\mathrm{self}}$. Note that ${%
\mathcal{F}}_{0}^{\mathrm{corr}}$ takes into account the fact that each ion
in the electrolyte is surrounded mostly by oppositely charged ions, which
amounts to the standard correlation free energy expression ${\mathcal{F}}%
_{0}^{\mathrm{corr}}=-\kappa _{0}^{3}/12\pi $, where $\kappa
_{0}=(8\pi \ell _{\mathrm{B}}^{0}c_{0})^{1/2}$ is the bulk inverse screening
length with $\ell _{\mathrm{B}}^{0}=e^{2}/(4\pi \varepsilon _{\mathrm{w}%
}\varepsilon _{0}kT)$ being the bulk Bjerrum length \footnote{Resibois, P. M. V. \emph{Electrolyte Theory}; Harper \& Row: New York, 1968.
}. It follows that within
the DH approximation, the mean-field free energy ${\mathcal{F}}_{0}^{\mathrm{%
MF}}$ is dominated by the entropy of the ions, which is well-approximated by
that of an ideal gas of particles, i.e. ${\mathcal{F}}_{0}^{\mathrm{MF}%
}\simeq 2(c_{0}\ln c_{0}-c_{0})$ assuming the bulk 1-1 electrolyte
concentration of $c_{0}$. Finally the self-energy contribution of ions reads 
${\mathcal{F}}_{0}^{\mathrm{self}}=2c_{0}\ell _{\mathrm{B}}^{0}/2a$ (in
units of $kT$).

Similar expressions may be written for the PEMs medium using the inverse
screening length $\kappa = (8\pi \ell_{\mathrm{B}} c)^{1/2}$, the Bjerrum
length $\ell_{\mathrm{B}}=e^2/(4\pi \varepsilon\varepsilon_0kT)$ and the
ionic concentration $c$. In this case, the mean-field free energy is again
approximated by that of an ideal gas since the net charge within the PEMs is
assumed to be zero. One thus has ${\mathcal{F}}_{\mathrm{P}}(c) = 2(c\ln c-
c) - \kappa_0^3/(12\pi) + 2c\ell_{\mathrm{B}}/2a$.

Thermodynamic equilibrium is imposed by minimizing ${\mathcal{F}}_{\mathrm{P}%
}(c)+{\mathcal{F}}_{0}(c_{0})$ with respect to the ion concentration $c$\
inside the PEMs, assuming that $c+c_{0}$ is constant, which is equivalent to
setting equal chemical potentials for the two media. One thus finds the
ion-partitioning law $c=c_{0}\exp (-\Delta \mu /kT)$, where 
\begin{equation*}
\Delta \mu =\frac{\ell _{\mathrm{B}}}{2}\left( \frac{1}{a}-\frac{1}{\kappa
^{-1}}\right) -\frac{\ell _{\mathrm{B}}^{0}}{2}\left( \frac{1}{a}-\frac{1}{%
\kappa _{0}^{-1}}\right) .
\end{equation*}%
The terms proportional to $1/a$ (with $a$ being the mean ion radius) are due
to the self-energy (Born energy) difference of the ions in the two media,
whereas the terms proportional to the Debye screening length come from the
DH correlation free energy difference. Since $\kappa ^{-1}$ and $\kappa
_{0}^{-1}$ are typically much larger than $a$, one can neglect this latter
contribution. In fact, an explicit estimate of the dielectric constant of
the PEMs, $\varepsilon$, including the DH correlation free energy shows
slightly larger values (up to a few percents) as compared with the results
reported in the text (obtained based only on the Born energy). The
difference is however within the experimental error bars.

\end{document}